\documentclass[letterpaper]{jpconf}
\usepackage{graphicx}
\usepackage{amsmath,amssymb,amsfonts,amsthm}
\usepackage[all,cmtip]{xy}

\catcode`,\active

\catcode`\,12

\newcommand*\pFqskip{8mu}
\catcode`,\active
\newcommand*\pFq{\begingroup
        \catcode`\,\active
        \def ,{\mskip\pFqskip\relax}%
        \dopFq
}
\catcode`\,12
\def\dopFq#1#2#3#4#5{%
        {}_{#1}F_{#2}\biggl[\genfrac..{0pt}{}{#3}{#4};#5\biggr]%
        \endgroup
}
\newcommand{\pd}{\partial}
\numberwithin{equation}{section}

\begin{document}
\title{The Racah algebra and superintegrable models}
\author{Vincent X. Genest}
\address{Centre de Recherches Math\'ematiques, Universit\'e de Montr\'eal, C.P. 6128, Centre-ville station, Montr\'eal (QC), Canada, H3C 3J7}
\ead{genestvi@crm.umontreal.ca}
\author{Luc Vinet}
\address{Centre de Recherches Math\'ematiques, Universit\'e de Montr\'eal, C.P. 6128, Centre-ville station, Montr\'eal (QC), Canada, H3C 3J7}
\ead{vinetl@crm.umontreal.ca}
\author{Alexei Zhedanov}
\address{Donetsk Institute for Physics and Technology, Donetsk 83114, Ukraine}
\ead{zhedanov@kinetic.ac.donetsk.ua}
\begin{abstract}
The universal character of the Racah algebra will be illustrated by showing that it is at the center of the relations between the Racah polynomials, the recoupling of three $\mathfrak{su}(1,1)$ representations and the symmetries of the generic second-order superintegrable model on the 2-sphere.
\end{abstract}
\section{Introduction}
This paper offers a review of the central role that the Racah algebra plays in connection with superintegrable models \cite{Genest-2013-tmp-1}.
\subsection{Superintegrable models}
A quantum system with $d$ degrees of freedom described by a Hamiltonian $H$ is maximally superintegrable (S.I.) if it possesses $2d-1$ algebraically independent constants of motion $S_i$ (also called symmetries) such that:
\begin{align}
[S_i,H]=0,\qquad 1\leq i\leq 2d-1,
\end{align}
where one of the symmetries is the Hamiltonian. Since the maximal number of symmetries that can be in involution is $d$, the constants of motion of a superintegrable system generate a non-Abelian algebra whose representations can in general be used to obtain an exact solution to the dynamical equations. A S.I. system is said to be of order $\ell$ if the maximal order of the symmetries in the momenta (apart from $H$) is $\ell$. We shall be concerned here with second-order ($\ell=2$) S.I. systems for which the Schr\"odinger equation is known to admit separation of variables and for which the symmetry algebras are quadratic.

S.I. systems, which include the classical examples of the isotropic harmonic oscillator and of the Coulomb-Kepler problem, are most interesting as models in applications and for pedagogical purposes. In particular, they form the bedrock for the analysis of symmetries and their description. Their study has helped to understand how Lie algebras, superalgebras, quantum algebras, polynomials algebras and algebras with involutions serve that purpose.
\subsection{Second-order S.I. systems in 2D}
The model that we shall focus on is the generic 3-parameter system on the 2-sphere. Its Hamiltonian is
\begin{align}
\label{Hamiltonian}
H=J_1^2+J_2^2+J_3^2+\frac{a_1}{x_1^2}+\frac{a_2}{x_2^2}+\frac{a_3}{x_3^2},\qquad a_i=k_i^2-1/4,
\end{align}
where
\begin{align}
x_1^2+x_2^2+x_3^2=1,
\end{align}
and where
\begin{align}
\label{Def-J}
J_1=-i(x_2\pd_{x_3}-x_3\pd_{x_2}),\quad J_2=-i(x_3\pd_{x_1}-x_1\pd_{x_3}),\quad J_3=-i(x_1\pd_{x_2}-x_2\pd_{x_1}),
\end{align}
are the familiar angular momentum operators. This 3-parameter system is 2\textsuperscript{nd} order superintegrable \cite{Kalnins-1996}. Most importantly, all 2\textsuperscript{nd} S.I. systems in 2D can be obtained from this model through specializations, limits and contractions \cite{Kalnins-2013-05}. The superintegrability of $H$ is confirmed by checking that the two operators:
\begin{align}
\label{Constants}
L_1=J_1^2+\frac{a_2 x_3^2}{x_2^2}+\frac{a_3 x_2^2}{x_3^2},\quad L_2=J_2^2+\frac{a_3 x_1^2}{x_3^2}+\frac{a_1 x_3^2}{x_1^2},
\end{align}
commute with $H$. Kalnins, Miller and Pogosyan \cite{Kalnins-1996} have examined the symmetry algebra and presented it as follows. With
\begin{align}
\label{Constants-2}
L_3=J_3^2+\frac{a_1 x_2^2}{x_1^2}+\frac{a_2 x_1^2}{x_2^2},\qquad R=[L_1,L_2],
\end{align}
one has $[H,L_3]=0$, $H=L_1+L_2+L_3+a_1+a_2+a_3$,
and the relations (with $(ijk)$ cyclic)
\begin{subequations}
\begin{align}
[L_i,R]&=4\{L_i,L_j\}-4\{L_i,L_k\}-(8-16a_j)L_j+(8-16a_k)L_{k}+8(a_j-a_k),
\\
&\begin{aligned}
R^2=&-\frac{8}{3}\,\{L_1,L_2,L_3\}-\sum_{i=1}^{3}\left\{(12-16a_i)L_i^2+\frac{1}{3}(16-176a_i)L_i+\frac{32}{3}a_i\right\}\\
&+\frac{52}{3}\left(\{L_1,L_2\}+\{L_2,L_3\}+\{L_1,L_3\}\right)+48(a_1a_2+a_2a_3+a_3a_1)\\
&-64\,a_1a_2a_3,
\end{aligned}
\end{align}
\end{subequations}
where $\{A,B\}=AB+BA$. Remarkably, Kalnins, Miller and Post \cite{Kalnins-2007-09} have further shown that the quadratic symmmetry algebra can be realized in terms of the difference operators associated to the Racah polynomials, that these same polynomials occur as transition coefficients between bases in which $L_1$ or $L_2$ is diagonal and furthermore that contractions of representations of the symmetry algebra lead to the symmetry algebras of the other 2\textsuperscript{nd} order S.I. systems and other families of orthogonal polynomials \cite{Kalnins-2013-05}. Details about these Racah polynomials \cite{Koekoek-2010}, denoted by $R_{n}(\lambda(x);\alpha,\beta,\gamma,\delta)$, will be given later. Suffice it to say for now that they are defined in terms of generalized hypergeometric functions, that they are of degree $n$ in the variable $\lambda(x)=x(x+\gamma+\delta+1)$, that they obey a discrete/finite orthogonality relation and that they sit atop the discrete part of the Askey scheme of hypergeometric orthogonal polynomials \cite{Koekoek-2010}.
\subsection{Objectives}
In reduced form, the (quadratic) Racah algebra has three generators $K_1$, $K_2$, $K_3$ and the defining relations
\begin{align}
[K_1,K_2]&=K_3,\nonumber\\
[K_2,K_3]&=K_2^2+\{K_1,K_2\}+d K_2+e_1,\\
[K_3,K_1]&=K_1^2+\{K_1,K_2\}+d K_1+e_2,\nonumber
\end{align}
where $d$, $e_1$ and $e_2$ are real parameters. The objectives of this paper are to show that this algebra has a universal character and intimately connects the generic second-order S.I. system on $S^2$, the Racah polynomials and the recoupling of three $\mathfrak{su}(1,1)$ representations. Schematically, the goal is to explain the links represented on the following diagram:
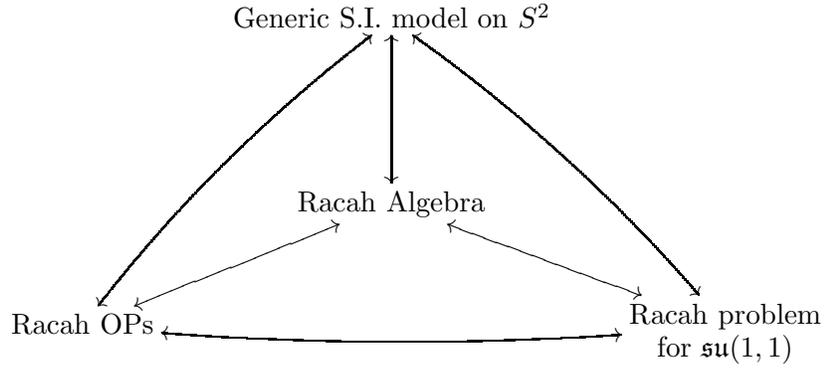
\begin{equation*}
\begin{aligned}
\xymatrix{
  & \text{Generic S.I. model on $S^{2}$} \ar@{<->}[dd] & 
\\
& &
\\
  & \text{Racah Algebra}\ar@{<->}[dl] \ar@{<->}[dr]& 
\\
\text{Racah OPs} \ar@/^/@{<->}[uuur]\ar@/_/@{<->}[rr] & &  \txt{Racah problem \\for $\mathfrak{su}(1,1)$} \ar@/_/@{<->}[uuul]
}
\end{aligned}
\end{equation*}
\section{Warming up with a simple model}
A key idea in our considerations is that 2\textsuperscript{nd} order superintegrable models can be obtained by combining 1D models that are exactly solvable. This can be done in simple cases by straightforward constructions \cite{Letourneau-1995-10} and has also been realized in the $R$-matrix formalism \cite{Harnad-2003,Kuznetsov-1992}.

Consider the 2D isotropic singular oscillator
\begin{align}
H=H_{x_1}+H_{x_2},
\end{align}
where
\begin{align}
H_{x_i}=-\frac{1}{2}\pd_{x_i}^2+\frac{1}{2}\left(x_i^2+\frac{a_i}{x_i^2}\right),\qquad a_i=k_i^2-1/4.
\end{align}
This is one of the four systems in the classification of second order S.I. systems in Euclidean space \cite{Winter-1965-06}. The spectrum of $H$ is
\begin{align}
E_{N}=N+(k_1+k_2+1)/2,\qquad N=n_{1}+n_{2},\qquad n_i\in \mathbb{N},
\end{align}
and has a $(N+1)$-fold degeneracy. It is well known that the associated Schr\"odinger equation separates in Cartesian and polar coordinates. To confirm that this system is maximally superintegrable, one needs to identify two independent constants of motion. This can be done using the $\mathfrak{su}(1,1)$ dynamical algebra of the one-dimensional components. Let
\begin{align}
B_{i}^{\pm}=\frac{1}{2}\left[(x_i\mp \pd_{x_i})^2-\frac{a_i}{x_i^2}\right],\quad i=1,2,
\end{align}
it is readily verified that these operators combine with $H_{x_i}$ to realize the $\mathfrak{su}(1,1)$ algebra since
\begin{align}
[J_0,J_{\pm}]=\pm J_{\pm},\qquad [J_{+},J_{-}]=-2J_{0},
\end{align}
with 
$$
J_0=H_{x_i}/2\qquad \text{and}\qquad J_{\pm}=B_{i}^{\pm}/2.
$$
In the positive-discrete series of $\mathfrak{su}(1,1)$ representations thus constructed, $B_{i}^{\pm}$ act as raising and lowering operators with respect to the eigenvalues of $H_{x_i}$ labeled by $n_{i}$. The conserved quantities of the 2D Hamiltonian $H=H_{x_1}+H_{x_2}$ are readily obtained from this observation. Indeed, combining the raising operator for $H_{x_1}$ with the lowering operator for $H_{x_2}$ (or vice-versa) will give an operator that leaves the total energy $E_{N}$ unchanged and that commutes with $H$. This is a straightforward generalization of the Schwinger construction for the isotropic harmonic oscillator. The superintegrability of $H$ is thus made manifest by exhibiting the operators
\begin{align}
C^{+}=B_{x_1}^{+}B_{x_2}^{-},\qquad C^{-}=B_{x_1}^{-}B_{x_2}^{+},
\end{align}
to which we conveniently add
\begin{align}
D=H_{x_1}-H_{x_2},
\end{align}
and by noting that
\begin{align}
[H,C^{\pm}]=0,\qquad [H,D]=0.
\end{align}
Defining relations for the symmetry algebra formed by the operators $C^{\pm}$ and $D$ are straightforwardly obtained \cite{Letourneau-1995-10}:
\begin{align}
\begin{aligned}
\label{Sym-Al}
[D,C^{\pm}]&=\pm 4 C^{\pm},
\\
[C^{-},C^{+}]&=D^{3}+\alpha_1 D+\alpha_2,
\end{aligned}
\end{align}
where
\begin{align}
\alpha_1=-H^2-2(k_1^2+k_2^2-2),\qquad \alpha_2=(2k_1^2-2k_2^2)H.
\end{align}
Since $H$ is central, $\alpha_1$ and $\alpha_2$ can be treated as constants on eigenspaces of $H$. An algebraic solution of the problem is obtained by working out the appropriate representations of this algebra. A special case of \eqref{Sym-Al} was found by Higgs in \cite{Higgs-1979} as symmetry algebra of the Coulomb problem on $S^2$. By taking a different set of operators, it is possible to cast the relations \eqref{Sym-Al} in a form that we would say is standard. Let
\begin{align}
K_1=\frac{1}{8}\Big(H_{x_1}-H_{x_2}\Big),\qquad K_2=\frac{1}{8}\left(C^{+}+C^{-}+\frac{1}{2}(D^2-H^2)\right).
\end{align}
It is seen that $K_2$ can be written as
\begin{align}
K_2=\frac{1}{8}\left((x_1\pd_{x_2}-x_2\pd_{x_1})^2-\frac{a_1 x_2^2}{x_1^2}-\frac{a_2 x_1^2}{x_2^2}-1/2\right),
\end{align}
and is purely angular in polar coordinates. If we take 
\begin{align}
K_3=[K_1,K_2]=\frac{1}{16}(C^{+}-C^{-}),
\end{align}
the defining relations become
\begin{align}
\begin{aligned}
\phantom{}
[K_1,K_2]&=K_3,
\\
[K_2,K_3]&=\{K_1,K_2\}+\delta_1 K_1+\delta_2,
\\
[K_3,K_1]&= K_1^2-\frac{1}{4}K_2+\delta_{3},
\end{aligned}
\end{align}
with
\begin{align}
\delta_{1}=-\frac{1}{4}(k_1^2+k_2^2-2),\quad \delta_{2}=\frac{1}{32}(k_1-k_2)(k_1+k_2)H,\quad \delta_{3}=-\frac{1}{64}H^2.
\end{align}
This presentation allows the identification with the Hahn algebra, a special case of the ``generic" Racah algebra (see next section). It is known to appear in connection with the Clebsch-Gordan problem for $\mathfrak{su}(1,1)$ \cite{Zhedanov-1993}. This suggests a potential link between the isotropic singular oscillator in two dimensions and the Clebsch-Gordan problem for the dynamical algebra of its one-dimensional components. We shall now proceed to discuss the most general second order S.I. model in two dimensions along the lines followed in this section and shall see that the link mentioned above is not fortuitous. However before we do so, we shall introduce thoroughly the Racah algebra, present its finite-dimensional representations and go over the relations these have with Racah polynomials.
\section{The Racah algebra}
The Racah algebra has three generators $K_1$, $K_2$ and $K_3$. In the generic presentation, they obey the relations
\begin{align}
\begin{aligned}
\phantom{}
\label{Racah}
[K_1,K_2]&=K_3,
\\
[K_2,K_3]&=a_2 K_2^2+a_1 \{K_1,K_2\}+c_1 K_1+d K_2+e_1,
\\
[K_3,K_1]&=a_1 K_1^2+a_2 \{K_1,K_2\}+c_2 K_2+ d K_1+e_2,
\end{aligned}
\end{align}
where the parameters $a_1$, $a_2$, $c_1$, $c_2$, $d$, $e_1$ and $e_2$ are taken to be real. This defines the most general associative quadratic algebra with two independent generators and a ladder property. To see this, let $K_1$, $K_2$ be the two independent generators and define $[K_1,K_2]=K_3$. $K_1$ and $K_2$ are assumed to be Hermitian, $K_3$ is thus anti-Hermitian. Consider the most general quadratic relations compatible with the hermiticity conditions:
\begin{align}
\begin{aligned}
\phantom{}
[K_2,K_3]&=a_2 K_2^2+a_1 \{K_1,K_2\}+g_1 K_1^2+h_1 K_3^2+c_1 K_1+d_1 K_2+e_1,
\\
[K_3,K_1]&=a_3 K_1^2+a_4 \{K_1,K_2\}+g_2 K_2^2+h_2 K_3^2+c_2 K_2+d_2 K_1+e_2.
\end{aligned}
\end{align}
It follows from the Jacobi identity 
$$[K_1,[K_2,K_3]]+[K_3,[K_1,K_2]]+[K_2,[K_3,K_1]]=0,
$$
that
\begin{align}
d_1=d_2,\quad a_3=a_1,\quad a_4=a_2,\quad h_1=h_2=0.
\end{align}
One then requires $g_1=g_2=0$ to ensure the ladder property (see later) and thus recovers \eqref{Racah}. This algebra made its appearance in the work of Granovskii and Zhedanov \cite{Granovskii-1988-10} where it was used in the context of the Racah problem of $\mathfrak{su}(2)$ to derive the symmetry group of the $6j$-symbols. It is also known as the Racah-Wilson algebra. When neither $a_1$ nor $a_2$ are zero, that is when $a_1\cdot a_2\neq 0$, the relations can be put in the following canonical form
\begin{subequations}
\label{Racah-2}
\begin{align}
\label{A}
[K_1,K_2]&=K_3,
\\
\label{B}
[K_2,K_3]&=K_2^2+\{K_1,K_2\}+d K_2+e_1,
\\
\label{C}
[K_3,K_1]&=K_1^2+\{K_1,K_2\}+d K_1+e_2,
\end{align}
\end{subequations}
where $d$, $e_1$ and $e_2$ are still real. This presentation thus retains three essential structure parameters and is arrived at by simple affine transformations of the generators $K_i\rightarrow u_i K_i+v_i$, $i=1,2,3$. It is verified that this algebra has the following Casimir operator (central element)
\begin{align}
\begin{aligned}
\label{Cas}
Q=&\{K_1^2,K_2\}+\{K_1,K_2^2\}+K_1^2+K_2^2+K_3^2
\\
&+(d+1)\{K_1,K_2\}+(2e_1+d)K_1+(2e_2+d)K_2,
\end{aligned}
\end{align}
which is cubic in the generators and commutes with each one of them.
\section{Representations of the Racah algebra and Racah polynomials}
We now wish to point out the connection between the Racah algebra and the Racah orthogonal polynomials. This can be done in at least two ways:
\begin{enumerate}
\item By constructing the finite-dimensional representations of the algebra.
\item By realizing the algebra in terms of the operators associated to the polynomials.
\end{enumerate}
We shall describe these two approaches in the following. We shall begin this section though by registering the basic definitions and properties of the Racah polynomials that we shall use.
\subsection{Racah polynomials}
The Racah polynomials $R_{n}(\lambda(x);\alpha,\beta,\gamma,\delta)$ of degree $n$ in $\lambda(x)=x(x+\gamma+\delta+1)$ depend on four real parameters $\alpha,\,\beta,\,\gamma,\,\delta$ and are defined by the following explicit expression ($n\in \mathbb{N}$):
\begin{align}
\label{Hyper}
R_{n}(\lambda(x))=\pFq{4}{3}{-n,n+\alpha+\beta+1,-x,x+\gamma+\delta+1}{\alpha+1,\beta+\delta+1,\gamma+1}{1},
\end{align}
where ${}_pF_{q}$ is the generalized hypergeometric series
\begin{align}
{}_pF_{q}\left[
\begin{matrix}
a_1 & \cdots & a_{p}
\\
b_1 & \cdots & b_{q}
\end{matrix};
z
\right]
=\sum_{j=0}^{\infty}\frac{(a_1)_{j}\cdots (a_{p})_{j}}{(b_1)_{j}\cdots (b_{q})_{j}}\frac{z^{j}}{j!},
\end{align}
and 
\begin{align*}
(a)_{j}=a(a+1)\cdots (a+j-1).
\end{align*}
The series in \eqref{Hyper} truncates since $(-n)_{j}=0$ for $j\geqslant n+1$. The polynomials thus defined satisfy $R_{0}(\lambda(x))=1$ and a three-term recurrence relation of the form \cite{Koekoek-2010}
\begin{align}
\lambda(x) R_{n}(\lambda(x))=A_{n}R_{n+1}(\lambda(x))-(A_{n}+C_{n})R_{n}(\lambda(x))+C_{n}R_{n-1}(\lambda(x)),
\end{align}
with 
\begin{align}
\begin{aligned}
\label{eq7}
A_{n}&=\frac{(n+\alpha+1)(n+\alpha+\beta+1)(n+\beta+\delta+1)(n+\gamma+1)}{(2n+\alpha+\beta+1)(2n+\alpha+\beta+2)},
\\
C_{n}&=\frac{n(n+\alpha+\beta-\gamma)(n+\alpha-\delta)(n+\beta)}{(2n+\alpha+\beta)(2n+\alpha+\beta+1)}.
\end{aligned}
\end{align}
As usual it is assumed that $R_{-1}(\lambda(x))=0$. Like all polynomials of the Askey scheme, the Racah polynomials are bispectral: in addition to obeying the above recurrence relation, they are aso eigenfunctions of the difference equation
\begin{align}
\mathcal{L}\,R_{n}(\lambda(x))=n(n+\alpha+\beta+1)R_{n}(\lambda(x)),
\end{align}
where
\begin{align}
\mathcal{L}=B(x) T^{+}+D(x) T^{-}-(B(x)+D(x))\mathbb{I},
\end{align}
with 
\begin{align}
T^{\pm}f(x)=f(x\pm 1),
\end{align}
and
\begin{align}
\begin{aligned}
\label{eq9}
B(x)&=\frac{(x+\alpha+1)(x+\beta+\delta+1)(x+\gamma+1)(x+\gamma+\delta+1)}{(2x+\gamma+\delta+1)(2x+\gamma+\delta+2)},
\\
D(x)&=\frac{x(x-\alpha+\gamma+\delta)(x-\beta+\gamma)(x+\delta)}{(2x+\gamma+\delta)(2x+\gamma+\delta+1)}.
\end{aligned}
\end{align}
Provided one of the following truncation conditions holds:
\begin{align*}
\alpha+1=-N,\quad \beta+\delta+1=-N,\quad \gamma+1=-N,
\end{align*}
the Racah polynomials $R_{n}(\lambda(x))$ enjoy a finite orthogonality relation of the form
\begin{align}
\sum_{x=0}^{N}w_{x}R_{n}(\lambda(x))R_{m}(\lambda(x))=h_{n}\delta_{nm},
\end{align}
where $w_{x}$ and $h_{n}$ are known explicitly. (For more details on the Racah polynomials see \cite{Koekoek-2010} where in particular the limit relations to other OPs of the Askey scheme are provided.)
\subsection{Finite-dimensional representations}
We shall now describe the finite-dimensional unitary representations of the Racah algebra and sketch how they are obtained. We take the defining relations to be in the canonical form \eqref{Racah-2}. We begin by taking one generator, say $K_1$, to be diagnal on the representation space and proceed to show that the Racah algebra has a ladder property.

Let $\omega_p$ be a vector of the representation space such that
\begin{align}
\label{Eigen}
K_1 \omega_{p}=\lambda_{p}\omega_{p},\qquad p\in \mathbb{R}.
\end{align}
Suppose we look for another eigenvector $\omega_{p'}$ with eigenvalue $\lambda_{p'}$ that has the form
\begin{align}
\omega_{p'}=\left\{\alpha(p) K_1+\beta(p) K_2+\gamma(p) K_3\right\}\omega_{p},
\end{align}
where $\alpha(p)$, $\beta(p)$ and $\gamma(p)$ are coefficients. Imposing the eigenvalue equation
\begin{align}
K_1\omega_{p'}=\lambda_{p'}\omega_{p'},
\end{align}
using \eqref{Eigen} and the commutation relations \eqref{A} and \eqref{C}, it is seen that the eigenvalues $\lambda_{p'}$ must satisfy
\begin{align}
(\lambda_{p'}-\lambda_{p})^2+(\lambda_{p'}+\lambda_{p})=0.
\end{align}
For a given $\lambda_{p}$, there are two solutions which we can choose to call $\lambda_{p+1}$ and $\lambda_{p-1}$. Assuming that $\lambda_{p}$ is non-degenerate and denoting by $E_{\lambda_p}$ the one-dimensional eigenspace, it follows from the above considerations that a generic element of the algebra will map $E_{\lambda_{p}}$ onto $E_{\lambda_{p-1}}\oplus E_{\lambda_{p}}\oplus E_{\lambda_{p+1}}$. We can thus write 
\begin{align}
\begin{aligned}
\label{actions}
K_1\omega_p&=\lambda_p \omega_p,
\\
K_2 \omega_p&=U_{p+1}\omega_{p_1}+V_p \omega_{p}+U_{p}\omega_{p-1},
\\
K_3\omega_p&=[K_1,K_2]\omega_p=U_{p+1}g_{p+1}\,\omega_{p+1}-U_{p}g_{p}\,\omega_{p-1},
\end{aligned}
\end{align}
with
\begin{align}
g_{p}=\lambda_{p}-\lambda_{p-1},
\end{align}
observing that $K_2$ is tridiagonal and $K_3$ bidiagonal. Note that $K_2$ is self-adjoint if $U_p$ is real. Having understood that the representations of the Racah algebra have a ladder structure, we now wish to focus on those representations that are finite-dimensional and for which the spectrum of $K_1$ is discrete. In the following we shall hence replace the vectors $\omega_p$, $p\in \mathbb{R}$, by the vectors $\psi_{n}$, $n\in \mathbb{Z}$, labeled by the discrete index $n$. The actions \eqref{actions} become
\begin{align}
\begin{aligned}
\label{actions-2}
K_1\psi_n&=\lambda_{n}\psi_{n},
\\
K_2\psi_{n}&=U_{n+1}\psi_{n+1}+V_{n}\psi_{n}+U_{n}\psi_{n-1},
\\
K_3\psi_{n}&=U_{n+1}g_{n+1}\,\psi_{n+1}- U_{n}g_{n}\,\psi_{n-1}.
\end{aligned}
\end{align}
with 
\begin{align}
g_{n}=\lambda_{n}-\lambda_{n-1},
\end{align}
and there remains to determine $\lambda_{n}$, $V_{n}$ and $U_{n}$. From \eqref{C}, one finds
\begin{align}
\label{eq2}
\lambda_{n}=(\sigma-n)(n-\sigma+1)/2,\quad g_{n}=\sigma-n,
\end{align}
and
\begin{align}
V_{n}=-\frac{\lambda_n^2+d \lambda_n+e_2}{\lambda_n},
\end{align}
where $\sigma$ is an arbitrary real parameter. We observe that the spectrum is quadratic in $n$. To find $U_{n}$, one uses \eqref{B} that yields the following recurrence relation for $U_{n}^2$:
\begin{align}
\label{eq}
2(g_{n+3/2}U_{n+1}^2-g_{n-1/2}U_{n}^2)=V_{n}^2+(2\lambda_n+d)V_{n}+e_1.
\end{align}
Instead of solving directly \eqref{eq}, it is simpler to use the fact that the Casimir operator $Q$ given in \eqref{Cas} is constant on irreducible representation spaces in order to find an expression for 
$$
2(g_{n+3/2}g_{n}U_{n+1}^2+g_{n+1}g_{n-1/2}U_{n}^2).
$$
Upon eliminating $U_{n+1}$ with the help of this result and solving \eqref{eq} for $U_{n}^2$ one arrives at
\begin{align}
U_{n}^2=\frac{\mathcal{P}(g_{n}^2)}{64g_{n}^2g_{n-1/2}g_{n+1/2}},
\end{align}
where $\mathcal{P}(z)$ is the fourth degree polynomial
\begin{align}
\begin{aligned}
\mathcal{P}(z)=&z^4-(4d+2)z^3+(4d^2+4d+1+8 e_2-16 e_1)z^2
\\
&\qquad -4(d^2+2e_2+4de_2+4q)z+16 e_2^2.
\end{aligned}
\end{align}
In terms of the roots $\xi_k^2$ of $\mathcal{P}(g_n^2)$, we can write
\begin{align}
U_{n}^2=\frac{\prod_{k=1}^{4}(g_{n}^2-\xi_k^2)}{64 g_{n}^2g_{n-1/2}g_{n+1/2}}.
\end{align}
For finite-dimensional representations, the index $n$ is comprised in a finite interval $N_1\leq n \leq N_2$, $N_1,N_2\in \mathbb{Z}$ and the eigenvalues of $K_1$ are $\lambda_{N_1},\lambda_{N_1+1},\ldots,\lambda_{N_2}$. Setting the arbitrary parameter $\sigma$ in \eqref{eq2} equal to $N_1+\rho$ and calling $N=N_2-N_1$, we can equivalently restrict the label $n$ to be in the interval
\begin{align}
0\leq \sigma \leq N,
\end{align}
with the eigenvalues $\lambda_0, \,\lambda_1,\ldots,\lambda_N$ given by
\begin{align}
\lambda_n=(\rho-n)(n-\rho+1)/2.
\end{align}
Clearly, in a $(N+1)$-dimensional representation we must have $U_0=U_{N+1}=0$ so that $\psi_{-1}$ and $\psi_{N+1}$ cannot be reached by the actions of $K_2$ on $\psi_0$ and $\psi_{N}$, respectively. This will be observed if one of the zeros $\xi_k^2$, $k=1,2,3,4$, say $\xi_i^2$ is such that
\begin{subequations}
\label{eq56}
\begin{align}
\xi_i^2=g_0^2=\rho^2,
\end{align}
and another say $\xi_j^2$, verify
\begin{align}
\xi_{j}^2=g_{N+1}^2=(\rho-N-1)^2.
\end{align}
\end{subequations}
\subsection{Connection with Racah polynomials}
Having described the finite-dimensional representations of the Racah algebra in the eigenbasis of $K_1$, we might wonder what the picture is in the eigenbasis of $K_2$. It turns out to be very similar. Indeed, it is observed that the defining relations \eqref{Racah-2} are invariant under the exchanges
\begin{align}
K_1\leftrightarrow K_2,\qquad e_1\leftrightarrow e_2.
\end{align}
It follows that the representation in the bases $\{\phi_{s}\}$ in which $K_2$ is diagonal
\begin{align}
\label{eq3}
K_2\phi_{s}=\mu_{s}\phi_{s},
\end{align}
can be obtained from those in which $K_1$ is diagonal using this symmetry property. In this correspondence with the formulas of the last subsection, we make the replacement $n\rightarrow s$. It is clear that $\mu_s$ will have a from similar to $\lambda_n$ say
\begin{align}
\mu_s=(\nu-s)(\nu-s+1)/2,
\end{align}
and that $K_1$ will be tridiagonal in the $\phi_{s}$ basis. In other words, $K_1$ and $K_2$ realize a Leonard pair \cite{Terwilliger-2001-06}. We now claim that the Racah polynomials arise as overlaps between the $\{\psi_{n}\}$ and the $\{\phi_{s}\}$ bases; that is, the Racah polynomials appear as expansion coefficients between the basis in which $K_1$ is diagonal and the one in which $K_2$ is diagonal.

Let us make this more explicit. Since the bases $\{\phi_{s}\}$ and $\{\psi_{n}\}$ span isomorphic spaces we can expand the elements of one in terms of those of the other:
\begin{align}
\label{eq4}
\phi_{s}=\sum_{n=0}^{N}W_{n}(s)\psi_{n}.
\end{align}
Let us write the coefficients $W_{n}(s)$ in the form
\begin{align}
W_n(s)=w_{0}(s) P_{n}(\mu_s),
\end{align}
so that $P_0(\mu_s)=1$. Using \eqref{eq3} and \eqref{actions-2}, it is seen upon acting with $K_2$ on both sides of \eqref{eq4} that the quantities $P_{n}(\mu_s)$ obey
\begin{align}
\mu_s P_{n}(\mu_s)=U_{n+1} P_{n+1}(\mu_s)+V_{n} P_{n}(\mu_s)+U_{n}P_{n-1}(\mu_s).
\end{align}
Given the formulas for $U_{n}$ and $V_{n}$, this recurrence relation is seen to coincide with that of the Racah polynomials. The zeros $\xi_k^2$ are related to the parameters $\alpha,\beta,\gamma,\delta$ of the polynomials $R_{n}(\lambda(x);\alpha,\beta,\gamma,\delta)$. If the truncations conditions are satisfied through
\begin{align}
\xi_1=\rho,\qquad \xi_4=(\rho-N-1),
\end{align}
(recall \eqref{eq56}) the identification is achived by the following parametrization of the roots:
\begin{align}
\xi_1=-\frac{\alpha+\beta}{2},\quad \xi_2=\frac{\beta-\alpha}{2}+\delta,\quad \xi_3=\frac{\beta-\alpha}{2},\quad \xi_4=\gamma-\frac{\alpha+\beta}{2}.
\end{align}
\subsection{The Racah algebra from the Racah polynomials}
We have illustrated how the Racah polynomials can be obtained from the Racah algebra by constructing the finite-dimensional representations. Let us indicate now that conversely, the Racah algebra can be identified from the properties of the Racah polynomials. As shall be seen the Racah algebra encodes the bispectrality properties of the polynomials. These properties amount to the fact that in addition to satisfying a three-term recurrence relation (as all orthogonal polynomials must do), the Racah polynomials also obey a difference equation. Let us recall that these relations can be put in the form
\begin{subequations}
\label{eq10}
\begin{align}
x(x+\gamma+\delta+1)R_{n}(\lambda(x))=\mathcal{M} R_{n}(\lambda(x)),
\\
\mathcal{L}R_{n}(\lambda(x))=n(n+\alpha+\beta+1)R_{n}(\lambda(x)),
\end{align}
\end{subequations}
where the \emph{recurrence} operator $\mathcal{M}$ and the \emph{difference} operator $\mathcal{L}$ are given by
\begin{align}
\mathcal{M}&=A_{n} T_{n}^{+}+C_{n}T_{n}^{-}-(A_{n}+C_{n})\mathbb{I},
\\
\mathcal{L}&=B(x) T_{x}^{+}+D(x) T_{x}^{-}-(B(x)+D(x))\mathbb{I},
\end{align}
with $T_{n}^{\pm}f(n)=f(n\pm 1)$, $T_x^{\pm}f(x)=f(x\pm 1)$ and where $A_n$, $C_n$ are given in \eqref{eq7} and $B(x)$, $D(x)$ provided by \eqref{eq9}. Consider now the realization on functions of $x$, where $\widetilde{K}_1$ and $\widetilde{K}_2$ are the two operators occurring on the left-hand side of \eqref{eq10}, i.e. $\widetilde{K}_1$ is the recurrence operator and $\widetilde{K}_2$ is the difference operator:
\begin{align}
\widetilde{K}_1=x(x+\gamma+\delta+1),\qquad \widetilde{K}_2=\mathcal{L}.
\end{align}
Performing the affine transformations 
\begin{align}
K_1=u_1 \widetilde{K}_1+v_1,\qquad K_2=u_2 \widetilde{K}_2+v_2,
\end{align}
one finds that $K_1$, $K_2$ verify the Racah algebra relations \eqref{Racah-2} with
\begin{align*}
\begin{aligned}
e_1&=\frac{1}{4}
\left(\frac{\alpha-\beta}{2}\right)
\left(\frac{\alpha+\beta}{2}\right)
\left(\frac{\alpha+\beta}{2}-\gamma\right)
\left(\frac{\alpha-\beta}{2}-\delta\right),
\\
e_2&=\frac{1}{4}
\left(\frac{\gamma-\delta}{2}\right)\left(\frac{\gamma+\delta}{2}\right)
\left(\frac{\gamma+\delta}{2}-\alpha\right)
\left(\frac{\gamma-\delta}{2}-\beta\right),
\\
d&=\frac{1}{4}
\left\{
\left(\frac{\gamma-\delta}{2}\right)^2
+\left(\frac{\gamma+\delta}{2}\right)^2
+\left(\frac{\gamma+\delta}{2}-\alpha\right)^2
+\left(\frac{\gamma-\delta}{2}-\beta\right)^2-2
\right\}.
\end{aligned}
\end{align*}
We could obviously have taken the realization on functions of $n$ with
\begin{align}
\widehat{K}_1=\mathcal{M},\qquad \widehat{K}_2=n(n+\alpha+\beta+1),
\end{align}
which is bound to lead to the same algebra. The duality property of the Racah polynomials under the exchanges
\begin{align}
x\leftrightarrow n,\qquad \alpha\leftrightarrow \gamma,\qquad \beta\leftrightarrow \delta,
\end{align}
which follows from \eqref{eq10} is immediately seen to correspond to the symmetry of the Racah algebra under
\begin{align}
K_1\leftrightarrow K_2,\qquad K_3\leftrightarrow -K_3,\qquad e_1\leftrightarrow e_2,
\end{align}
that we already observed.
\section{The generic superintegrable model on $S^2$ and $\mathfrak{su}(1,1)$}
We now return to the generic superintegrable model on $S^2$ with Hamiltonian \eqref{Hamiltonian} and constants of motion \eqref{Constants} and \eqref{Constants-2}. We want to show that it is intimately connected to the $\mathfrak{su}(1,1)$ algebra.

Consider three $\mathfrak{su}(1,1)$ realizations identical to the one introduced in the discussion of the two-dimensional singular oscillator in Section 2:
\begin{align}
\label{rea}
J_0^{(i)}=\frac{1}{4}\left(-\pd_{x_i}^2+x_2^2+\frac{a_i}{x_i^2}\right),\quad J_{\pm}^{(i)}=\frac{1}{4}\left((x_i\mp \pd_{x_i})^2-\frac{a_i}{x_i^2}\right),
\end{align}
with $a_i=k_i^2-1/4$ and $i=1,2,3$. These provide positive discrete series representations for which the $\mathfrak{su}(1,1)$ Casimir element
\begin{align}
\label{dompe}
C^{(i)}=[J_0^{(i)}]^2-J_{+}^{(i)}J_{-}^{(i)}-J_0^{(i)},\quad i=1,2,3,
\end{align}
takes the value
\begin{align}
C^{(i)}=\nu_i(\nu_i-1),\qquad \nu_i=(k_i+1)/2.
\end{align}
These three sets of $\mathfrak{su}(1,1)$ generators can be added to produce a ``fourth'' realization:
\begin{align}
\label{algebra}
J_0^{(4)}=J_0^{(1)}+J_{0}^{(2)}+J_0^{(3)},\quad J_{\pm}^{(4)}=J_{\pm}^{(1)}+J_{\pm}^{(2)}+J_{\pm}^{(3)}.
\end{align}
Three types of Casimir operators can be distinguished in the process:
\begin{enumerate}
\item The ``initial'' Casimir operators \eqref{dompe}
\item The ``intermediate'' Casimir operators associated to the addition of two representations
\begin{align}
\label{inter}
C^{(ij)}=[J_0^{(i)}+J_{0}^{(j)}]^2-(J_{+}^{(i)}+J_{+}^{(j)})(J_{-}^{(i)}+J_{-}^{(j)})-(J_0^{(i)}+J_0^{(j)}),
\end{align}
with $(ij)=(12), (23), (31)$.
\item The ``full'' Casimir operator
\begin{align}
\label{full}
C^{(4)}=[J_0^{(4)}]^{2}-J_{+}^{(4)}J_{-}^{(4)}-J_{0}^{(4)}.
\end{align}
\end{enumerate}
By a direct computation one finds that
\begin{align}
C^{(ij)}=\frac{1}{4}\left\{J_k^2+\frac{a_i x_j^2}{x_i^2}+\frac{a_j x_i^2}{x_j^2}+a_i+a_j-1\right\},
\end{align}
with $(ijk)$ a cyclic permutation of $(1,2,3)$. Comparing with \eqref{Constants} and \eqref{Constants-2}, we see that
\begin{align}
L_1=4 C^{(23)}-a_2-a_3+1,\quad L_2= 4C^{(31)}-a_3-a_1+1,\quad L_3=4 C^{(12)}-a_1-a_2+1,
\end{align}
and thus observe that the constants of motion of the generic S.I. system on $S^2$ are basically the intermediate Casimir operators arising in the addition of three $\mathfrak{su}(1,1)$ representations. Similarly one can check that the full Casimir operator $C^{(4)}$ takes the following form when \eqref{rea} and \eqref{algebra} are used in \eqref{full}:
\begin{align}
C^{(4)}=\frac{1}{4}\left\{J_1^2+J_2^2+J_3^2+(x_1^2+x_2^2+x_3^2)\left(\frac{a_1}{x_1^2}+\frac{a_2}{x_2^2}+\frac{a_3}{x_3^2}\right)-\frac{3}{4}\right\}.
\end{align}
As a consequence,
\begin{align}
H=4 C^{(4)}+\frac{3}{4},
\end{align}
if $x_1^2+x_2^2+x_3^2=1$. At this point we may ask if this restriction to $S^2$ can generally be ensured in the addition of the three $\mathfrak{su}(1,1)$ representations. That the answer is yes is readily seen. It is noted from \eqref{rea} that
\begin{align}
2J_0^{(i)}+J_{+}^{(i)}+J_{-}^{(i)}=x_i^2,
\end{align}
and hence that
\begin{align}
S=2J_0^{(4)}+J_{+}^{(4)}+J_{-}^{(4)}=x_1^2+x_2^2+x_3^2.
\end{align}
Since $S$ commutes with $4C^{(4)}+3/4$, it is ``time-independent'' and as a constant can be taken to be 1. We can thus conclude the following. The generic S.I. 3-paramater system is obtained from the addition of three $\mathfrak{su}(1,1)$ realizations. In this identification the restriction $x_1^2+x_2^2+x_3^2=1$ to $S^2$ is preserved; the Hamiltonian corresponds to the full Casimir operator $C^{(4)}$ for the addition of the three representations and the constants of motion are obviously the intermediate Casimir operators $C^{(ij)}$ which commute with $C^{(4)}$. The algebra that these intermediate Casimir operators generate is thus the symmetry algebra of the generic $S.I.$ system on $S^2$. We shall show in the next section that the intermediate Casimir operators in the addition of three $\mathfrak{su}(1,1)$ irreducible representations generate the Racah algebra.

Note that the $R$-matrix approach has been used in \cite{Harnad-2002} to describe the generic 3-parameter model on $S^2$ and construct its invariants. One proceeds via dimensional reduction from $S^5$ to $S^2$ with a Lax matrix that also involves three $\mathfrak{su}(1,1)$ elements. It is interesting to further point out that the same generic S.I. model on $S^2$ has been shown to correspond to one of the Krall-Sheffer classes of orthogonal polynomials in two variables \cite{Harnad-2001,Vinet-2003}.
\section{The Racah problem for $\mathfrak{su}(1,1)$ and the Racah algebra}
We loosely refer to the Racah problem as the recoupling of three irreducible representations. Strictly speaking, it is about determining the unitary transformation between two canonical bases corresponding to the steps $(1\oplus 2)\oplus 3$ and $1\oplus (2\oplus 3)$ which are respectively associated to the diagonalization of the intermediate Casimir operators $C^{(12)}$ and $C^{(23)}$. Our goal in this section is to comment on the algebra that $C^{(12)}$ and $C^{(23)}$ generate in the case of $\mathfrak{su}(1,1)$ (and of $\mathfrak{su}(2)$ as a matter of fact).

Consider the addition of three $\mathfrak{su}(1,1)$ representations as in \eqref{algebra} and take each initial Casimir operator $C^{(i)}$ to be a multiple of the identity:
\begin{align}
C^{(i)}=\lambda_i,\qquad i=1,2,3.
\end{align}
The intermediate Casimir operators $C^{(12)}$ and $C^{(23)}$ (given by \eqref{inter}) can then be expressed as follows
\begin{align}
\begin{aligned}
C^{(12)}&=2J_0^{(1)}J_0^{(2)}-(J_{+}^{(1)}J_{-}^{(2)}+J_{-}^{(1)}J_{+}^{(2)})+\lambda_1+\lambda_2,
\\
C^{(23)}&=2J_0^{(2)}J_0^{(3)}-(J_{+}^{(2)}J_{-}^{(3)}+J_{-}^{(2)}J_{+}^{(3)})+\lambda_2+\lambda_3.
\end{aligned}
\end{align}
Further assume that the full Casimir operator $C^{(4)}$ which can be written in the form
\begin{align}
C^{(4)}=C^{(12)}+C^{(23)}+C^{(31)}-C^{(1)}-C^{(2)}-C^{(3)},
\end{align}
is also a multiple of the identity, i.e.
\begin{align}
C^{(4)}=\lambda_4.
\end{align}
Denoting by $V^{(\lambda_i)}$ an irreducible $\mathfrak{su}(1,1)$ representation space on which the Casimir operator $C^{(i)}$ is equal to $\lambda_i$, we are thus looking at the decomposition of $V^{(\lambda_1)}\otimes V^{(\lambda_2)}\otimes V^{(\lambda_3)}$ in irreducible components $V^{(\lambda_4)}$. Consider now the algebra generated by the intermediate Casimir operators in this Racah problem. Define
\begin{align}
\kappa_1=-C^{(12)}/2\qquad \kappa_2=-C^{(23)}/2,
\end{align}
and let 
\begin{align}
\kappa_3=[\kappa_1,\kappa_2].
\end{align}
A direct computation in which $C^{(i)}$ is replaced by $\lambda_i$ gives, remarkably, the defining relations of the Racah algebra
\begin{align}
\begin{aligned}
\phantom{}
[\kappa_1,\kappa_2]&=\kappa_3,\\
[\kappa_2,\kappa_3]&=\kappa_2^2+\{\kappa_1,\kappa_2\}+d \kappa_2+e_1,\\
[\kappa_3,\kappa_1]&=\kappa_1^2+\{\kappa_1,\kappa_2\}+d \kappa_1+e_2,
\end{aligned}
\end{align}
where
\begin{align}
d=\frac{1}{2}(\lambda_1+\lambda_2+\lambda_3+\lambda_4)\quad e_1=\frac{1}{4}(\lambda_1-\lambda_4)(\lambda_2-\lambda_3)\quad e_2=\frac{1}{4}(\lambda_1-\lambda_2)(\lambda_4-\lambda_3)
\end{align}
The intermediate Casimir operators in the addition of three $\mathfrak{su}(1,1)$ representations form the Racah algebra which is thus the structure behind the Racah problem for $\mathfrak{su}(1,1)$. Combining this with the identification of the intermediate Casimir operators with the constants of motion of the generic S.I. 3-parameter system on $S^2$, it follows that the (reduced) Racah algebra is the symmetry algebra of this model.

We recall from Section (4.3) that the Racah polynomials appear as expansion coefficients between bases for Racah algebra representation spaces in which $K_1$ is diagonal on the one hand and $K_2$ is diagonal on the other. Since as we just have seen, $K_1$ and $K_2$ can be realized by $C^{(12)}$ and $C^{(23)}$, this naturally relates to the fact that the Racah coefficients (the elements of the matrix relating the bases associated to the 2 step-wise recoupling processes) are Racah polynomials. In the context of the superintegrable model on $S^2$, these Racah coefficients can be connected to separation of variables. This is seen as follows. The diagonalization of
\begin{align}
C^{(12)}=\frac{1}{4}(L_3+a_1+a_2-1)=J_3^2+\frac{a_1x_2^2}{x_1^2}+\frac{a_2 x_1^2}{x_2^2},
\end{align}
brings the separation of variables in the usual spherical coordinates (in which $x_1$ and $x_2$ are paired):
\begin{align}
\label{first}
x_1=r\sin \theta\cos \phi,\quad x_2=r\sin\theta\sin\phi,\quad x_3=r\cos \theta,
\end{align}
while the diagonalization of
\begin{align}
C^{(23)}=\frac{1}{4}(L_1+a_2+a_3-1)=J_1^2+\frac{a_2 x_3^2}{x_2^2}+\frac{a_3 x_2^2}{x_3^2},
\end{align}
leads to the separation of variables in another spherical coordinate system (in which $x_2$ and $x_3$ are paired) namely,
\begin{align}
\label{second}
x_1=r\cos\theta,\quad x_2=r\sin\theta\cos\phi,\quad x_3=r\sin\theta\sin\phi.
\end{align}
Hence the Racah polynomials are in this framework, the overlap coefficients between these two sets of wavefunctions of the generic 3-parameter system on $S^2$ which are obtained by separating the variables in the considered systems \eqref{first} and \eqref{second}.

Summing up, the identification of the generic S.I. model on $S^2$ as the full Casimir operator in the addition of three $\mathfrak{su}(1,1)$ realizations provides a natural way of obtaining the constants of motion (as the intermediate Casimir operators) and of determining the symmetry algebra (as the Racah algebra). This intimately associates the Racah polynomials to the model on $S^2$. It should be said that representations with continuous spectra are found to bring the Wilson polynomials in the picture. As explained in \cite{Kalnins-2013-05}, all the other second order S.I. mdels in two dimensions can be obtained from the generic system on $S^2$ by contractions and specializations. Correspondingly, when effected on the Racah algebra and the Racah/Wilson polynomials these operations provide the symmetry algebras of all these second order S.I. models and their tagging to orthogonal polynomials of the Askey scheme.
\section{Conclusion}
Let us summarize the main findings and offer perspectives. We presented the Racah algebra and its finite-dimensional representations. We proceeded to explain its universal character and the relations depicted on the diagram presented in Section I. It was shown that the Racah algebra is behind
\begin{itemize}
\item The generic 3-parameter superintegrable model on $S^2$ and hence all second order superintegrable systems in two dimensions
\item The Racah problem for $\mathfrak{su}(1,1)$, that is the combination of three irreducible $­\mathfrak{su}(1,1)$ representations
\item The Racah polynomials that sit atop the discrete part of the Askey scheme of hypergeometric orthogonal polynomials
\end{itemize}
Looking at three and higher dimensions, the analogous connections between superintegrable models, polynomial algebras and special functions are bound to be illuminating. They are expected to feed the theory of multivariate orthogonal polynomials and their algebraic interpretations.

As an illustration of this, let us mention that Kalkins, Miller and Post have already shown that 2-variable Racah-Wilson polynomials occur in the $S^3$ model \cite{Kalnins-2011-05}. The construction presented here extends to any dimensions and the three-dimensional model on $S^3$ arises from the addition of four $\mathfrak{su}(1,1)$ representations and corresponds to the $9j$ or Fano problem for $\mathfrak{su}(1,1)$. The symmetries in this case are expected to lead to a rank 2 version of the Racah algebra which should encode the properties of the general bivariate Racah-Wilson polynomials. We intend to pursue investigations along those lines.
\section*{Acknowledgements}
V.X.G. holds an Alexander-Graham-Bell fellowship from the Natural Sciences and Engineering Research Council of Canada (NSERC). The research of L.V. is supported in part by NSERC.
\section*{References}

\end{document}